# Predictive Neural Networks for Geophysics with Enhanced Seismic Imaging

Ping Lu[1], Yanyan Zhang[1], Jianxiong Chen[1], Yuan Xiao[1], George Zhao[2]

## Abstract

We propose a predictive neural network architecture that can be utilized to update reference velocity models as inputs to the full waveform inversion. Deep learning models are explored to augment velocity model building workflows during processing the 3D seismic volume in salt-prone environments. Specifically, a neural network architecture, with 3D convolutional, de-convolutional layers, and 3D max-pooling, is designed to take standard amplitude 3D seismic volumes as an input. Enhanced data augmentations through generative adversarial networks and a weighted loss function enable the network to train with few sparsely annotated slices. Batch normalization is also applied for faster convergence. A 3D probability cube for salt bodies and inclusions is generated through ensembles of predictions from multiple models in order to reduce variance. Velocity models inferred from the proposed networks provide opportunities for FWI forward models to converge faster with an initial condition closer to the true model. In addition, in each iteration step, the probability cubes of salt bodies and inclusions inferred from the proposed networks can be used as a regularization term within the FWI forward modelling, which may result in an improved velocity model estimation while the output of seismic migration can be utilized as an input of the 3D neural network for subsequent iterations.

## Introduction

Full waveform inversion (FWI) has become a popular method to estimate elastic earth properties from seismic data, and it has great utility in seismic velocity model building and seismic reflectivity imaging in areas of complex salt (Chen, et al., 2018) (Wang, Zhang, Mei, Lin, & Huang, 2019). FWI is a non-linear data-fitting procedure that matches the predicted to observed waveform data given an initial guess of the subsurface parameters. The velocity model parameters are updated to reduce the misfit between the observed and predicted data until the misfit is sufficiently small. Sharp velocity boundaries such as between salt and sediment are often updated manually for each iteration based on the seismic reflectivity images.

Deep learning is a new area of machine learning research. With massive amount of computations, it has an end-to-end learning structure, and replaces the traditional feature extraction pipeline in machine learning with a signal learning algorithm. Thanks to the success of computer vision, deep learning has been widely adopted in the geophysical interpretation and inversion workflows

[1]Anadarko Petroleum Corporation, The woodlands, TX, 77380
[2]WesternGeco, Houston, TX, 77042

(Lu, Morris, Brazell, Comiskey, & Xiao, 2018). A proposal to harness neural networks to assist the geophysical FWI workflows followed by current examples of their products and success is presented.

## Proposed Methods

Here, we introduce how to utilize the predicted probability cube inferred from a neural network based on seismic data to assist FWI workflow. The achieved salt mask probability cube is capable of assisting convergences of the velocity model in many aspects. In this paper, we discuss two major improvements contributed by the deep learning results.

1. Initialization

FWI is one of the techniques highly sensitive to its initial value of the velocity. Due to its non-linearity and non-convexity, a naïve guess may lead the results converge to a local-optima or saddle point, causing a geologically unacceptable velocity model. The salt body with inclusions probability learned by deep learning model, on the other hand, has a solid numerical interpretation on salt-dome especially for the salt boundary, which can be utilized to generate a reasonable inference.

A pre- or post-stack seismic volume, which is generated by 1-D depth-dependent sediment velocity profile, can be used as an input into the well-trained deep neural network to predict corresponding salt probability cube. However, unlike traditional method which treats the sediment velocity as the initial value and keeps updating on it, we use the predicted probability cube to create an initialization of the geological velocity model. Specifically, let $m_0$ denotes the initial guess for velocity model, $P_0$ as the first probability output of the deep learning network, we propose

$$m_0 = v_{sa} \cdot P_0 + v_{se} \cdot (1 - P_0),$$

where $v_{sa}$ is the empirical velocity value for salt dome, and $v_{se}$ is the depth-dependent sediment velocity profile.

2. Regularization

Mathematically, FWI process can be formulated as an optimization problem with object function

$$\min_{m} \Phi(m) = \sum_{i=1}^{n_s} \|F_i(m) - d_i\|^2.$$

Here inside the equation, function $F_i(\cdot)$ denotes the forward modeling operation, $d_i$ is considered as the observed pre-stack signal. In this problem, as we have achieved additional information for salt body probability cube in each iteration, it can be utilized as a priori information to the velocity model by adding a regularization term

$$\min_{m} \Phi'(m) = \sum_{i=1}^{n_s} \|F_i(m) - d_i\|^2 + \lambda \cdot \|m - R(P)\|^2,$$

such that $\lambda$ is an empirical regularization weight and $R(P) = v_{sa} \cdot P + v_{se} \cdot (1 - P)$ is considered as a prior function with deep learning probability output $P$. It can be seen that the initialization process is actually one realization of the prior function when $P = P_0$, however, as the initialization process is to provide a good starting point for velocity update, and the priori function $R(P)$ incorporated inside the regularization term pushes the predicted velocity model to match the probability information, it is better to distinguish the different objectives of these calculations.

As the FWI proceed, the velocity model keeps updating and ultimately it approximates the true velocity model. The convergence criteria in the problem is set as

$$\|\Phi'(m_{k+1}) - \Phi'(m_k)\| \leq \varepsilon,$$

where $\varepsilon$ is a pre-defined acceptable difference. It is also worth noting that, when the velocity converges, the predicted probability from deep learning (DL) network is close to a binary cube, such that

$$P_{k+1} \approx \begin{cases} 0 & \text{when the point is inside salt dome} \\ 1 & \text{when the point is out of salt dome} \end{cases}.$$

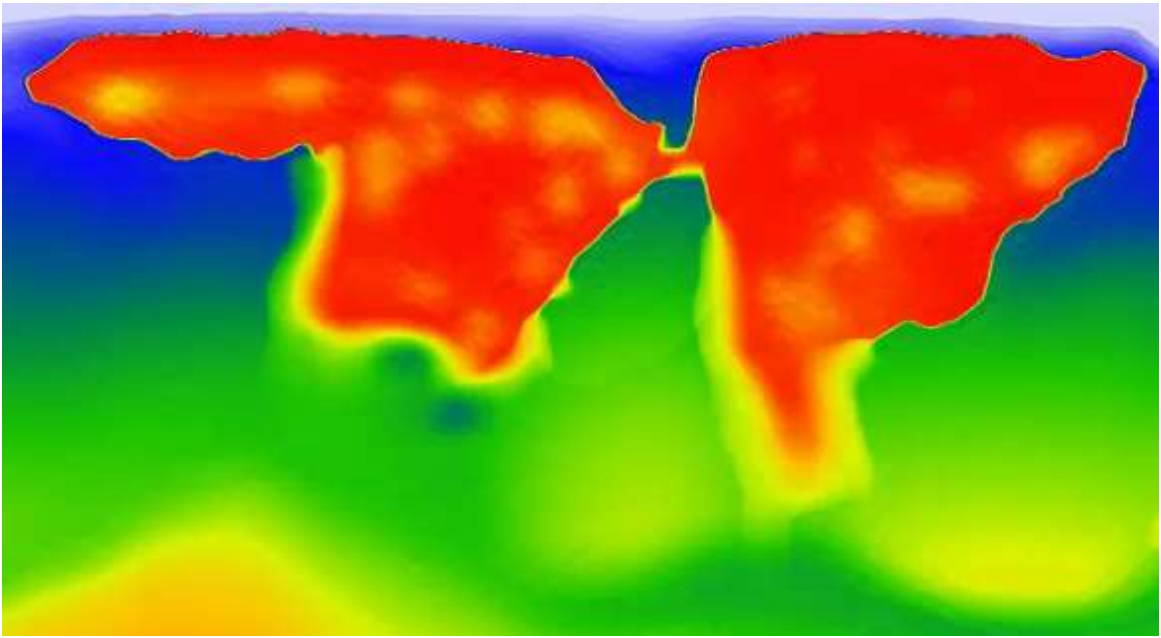

Figure 1. Velocity model from deep learning

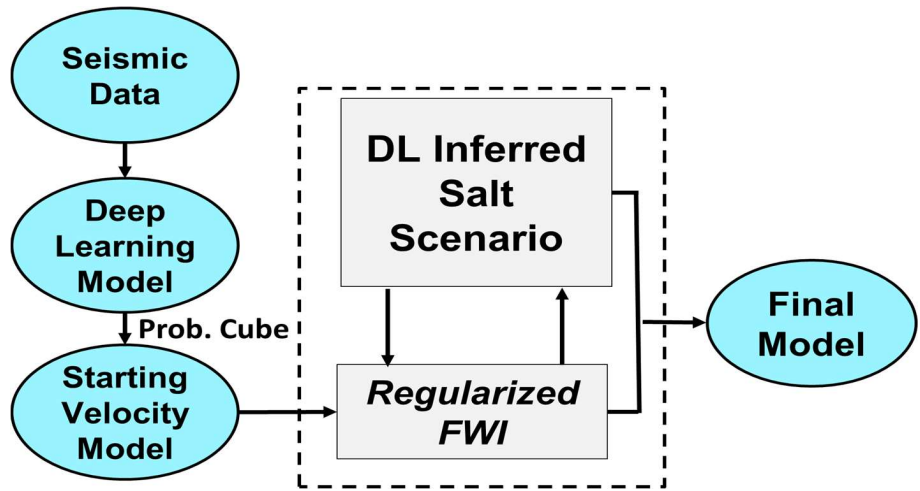

Chart 1. Workflow

Here, we present a flowchart to demonstrate the whole idea of how to integrate deep learning with FWI workflow.

**- Seismic Data** represents 3D pre- or post-stack seismic data.

**- Deep Learning Model** is used for detection of the most potential salt body including inclusions.

**- Starting Velocity Model (fig. 1)** represents the blended salt velocity converted from probability cube inferred by deep learning model before *FWI*.

**- DL Predicted Salt Scenario** is based on suggested velocity updates from regularized *FWI*.

**- Iterations** stops as convergence criteria satisfied.

**- Final Model** is different from the starting one in terms of salt velocity, salt geometry, and near-salt sediment velocities; and results in more sensible velocity model, flatter gathers, and more focused sub-salt image.

## Applications with a Field Example

The spatial consistency in salt body detection is barely addressed in general, while this is a remarkable aspect of human expertise. Our approach explicitly provides spatially consistent results as shown in fig.2.

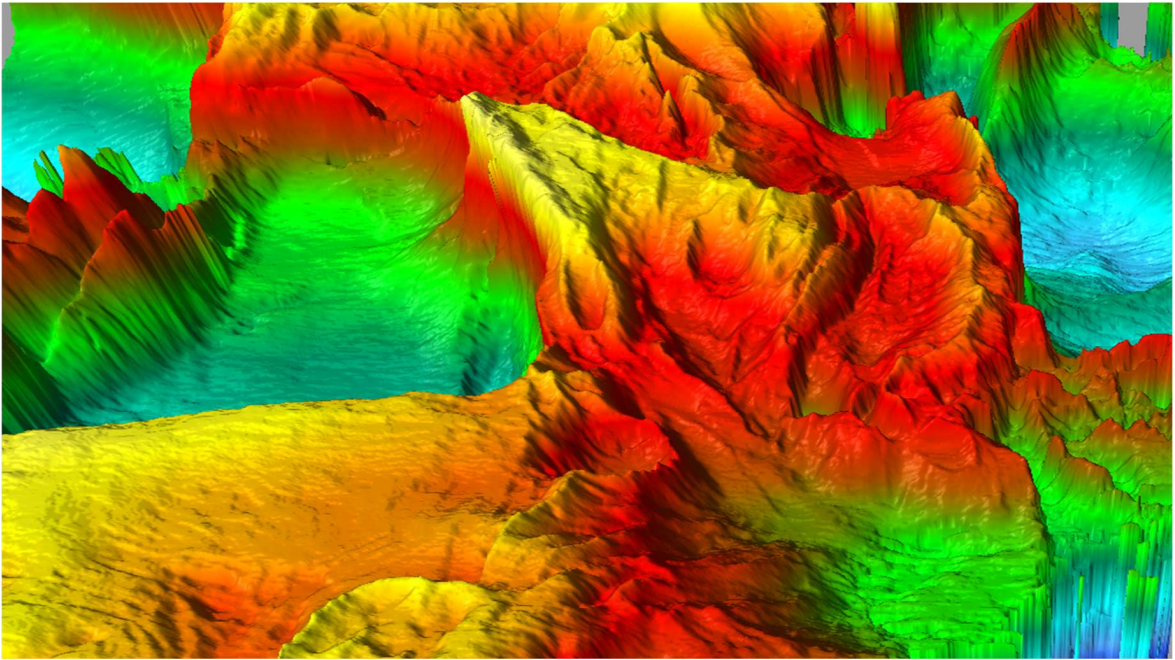

Figure 2. Visualization of 3D Salt Surface Inferred from DL

Altogether, experimental results generated from a real Gulf of Mexico seismic data suggest that the prediction represents a powerful framework, which provides complimentary information to the FWI procedure to generate a high resolution velocity model including an accurate salt model and ultimately a sharp subsalt image. Fig.3 (b) shows more continuous reflections versus fig.3 (a) in the subsalt region.

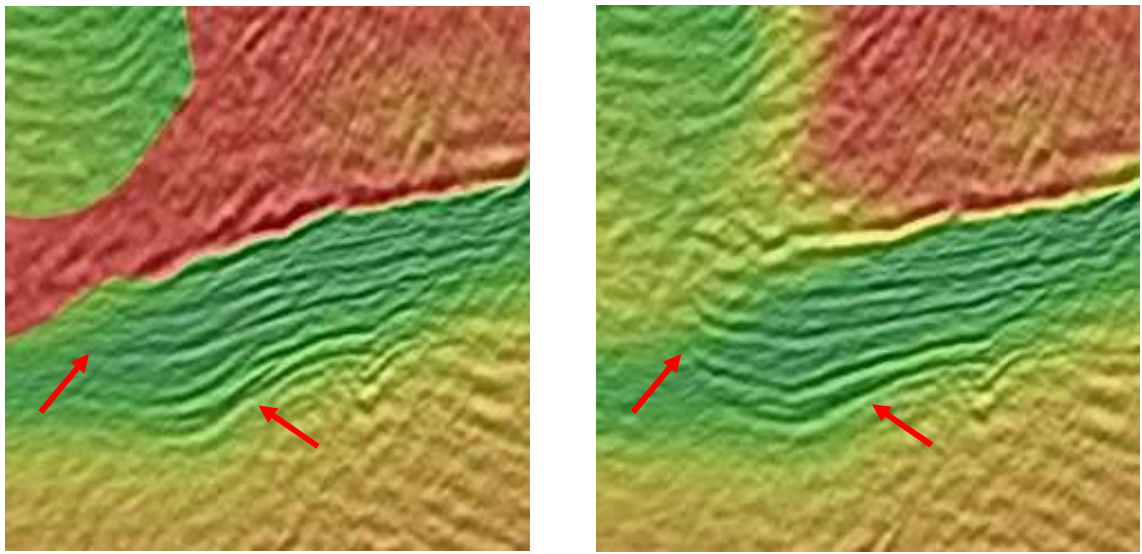

(a) (b)

Figure 3. Seismic imaging for subsalt

## Conclusion

A new workflow integrated with deep learning model is introduced for enhancing the seismic imaging quality with improved process of generation of salt velocity models. There are two significant benefits of leveraging the proposed approach compared with the conventional one. First, a salt velocity model, which could be used for initial velocity model for FWI algorithm, can be directly converted from a salt probability cube through inserting surrounding velocity model. A precise starting-velocity model is critical for the optimization process, which may end up finding the local minimum instead of the global minimum due to the initial guess far away from the ground truth. On the other hand, the salt probability cube could be used as an extra regularization term integrated in the objective function. Searching in three dimension in order to find out the optimal directions for optimization problems is an extremely challenging and time consuming task. In conjunction with output from deep learning model, the searching area is perfectly regularized in the salt probability mask region, which provides an opportunity to reach in convergence with a faster pace.

## Acknowledgement

The authors thank Anadarko for permission to publish this work.

## References

Chen, J., Sixta, D., Raney, G., Mount, V., Riddle, E., Nicholson, A., . . . Peng, C. (2018). Improved sub-salt imaging from reflection full waveform inversion guided salt scenario interpretation: A case history from deep water Gulf of Mexico. SEG Technical Program Expanded Abstracts, (pp. 3773-3777).

Lu, P., Morris, M., Brazell, S., Comiskey, C., & Xiao, Y. (2018). Using generative adversarial networks to improve deep-learning fault interpretation network. The Leading Edge, 578-583.

Wang, P., Zhang, Z., Mei, J., Lin, F., & Huang, R. (2019). Full-waveform inversion for salt: A coming of age . The Leading Edge, 204-213.